\documentclass[a4paper,11pt]{extarticle}
\usepackage{jheppub}

\usepackage{amsmath,amsfonts}
\usepackage{epsfig}

\makeatletter
\def\@fpheader{\relax}
\makeatother

\def\lam{\lambda}
\def\d{\partial}
\def\l{\left(}
\def\r{\right)}

\newcommand{\be}{\begin{equation}}
\newcommand{\ee}{\end{equation}}
\newcommand{\bg}{\begin{gather}}
\newcommand{\eg}{\end{gather}}
\newcommand{\bseq}{\begin{subequations}}
\newcommand{\eseq}{\end{subequations}}

\def\half{\frac{1}{2}}

\newcommand{\db}{\ensuremath{\delta\beta}}

\newcommand{\scA}{\ensuremath{\mathcal{A}}}
\newcommand{\scC}{\ensuremath{\mathcal{C}}}

\newcommand{\sph}{\ensuremath{\mathbf{S}^2}}

\title{On critical dimension in spherical black brane phase transition}

\author{Andrei Khmelnitsky}

\affiliation{Arnold Sommerfeld Center for Theoretical Physics 
Ludwig-Maximilians-Universit\"at M\"unchen,
80333 Munich, Germany}
\emailAdd{khmelnitskiy@physik.lmu.de}

\abstract{We study the Gregory-Laflamme instability of a large uniform black brane wrapping a two-sphere compactification manifold. This paper continues the work~\cite{Kol}, where the compactifications on $p$-torus were considered. The new features of the spherical case are the non-zero curvature of the compactification manifold and the absence of the rescaling symmetry due to a built-in stabilization mechanism. We calculate the order of the phase transition in dependence on the number $d$ of extended dimensions using the Landau-Ginzburg approach. It is found that for $d > 11$ a uniform spherical black brane in microcanonical ensemble exhibits a smooth second order phase transition towards a stable branch of non-uniform black brane solutions. The critical number of extended dimensions, for which there is a change in the order of the phase transition, is different for microcanonical and canonical ensembles and does not coincide with the critical number of dimensions in the case of the flat toric compactifications. We briefly discuss the origin of this mismatch in the orders of phase transition for the different ensembles.
\vspace{.3\textheight}}

\notoc

\begin{document}

\maketitle

\flushbottom

\section{Introduction and summary}

Uniform extended black branes in the presence of compact extra dimensions are unstable with respect to long wavelength perturbations if the black brane horizon size is substantially smaller than the size of the extra dimensions. This is known as the Gregory-Laflamme (GL) instability~\cite{Gregory:1993vy}. Therefore a large black brane should undergo a phase transition once its size becomes smaller than a certain critical size. By studying non-uniform perturbations on top of the critical black brane it is possible to find the order of this phase transition. It could be either a smooth second (or possibly higher) order transition when the black brane becomes slightly non-uniform in the compact dimension in a continuos way or a first order transition when below the critical size the black brane decays into some completely different solution. The final stage of the first order transition is in general unknown (see~\cite{Kol:2004ww,Harmark:2005pp,Niarchos:2008jc} for review).
The first study of this kind was performed by Gubser for a five-dimensional black string on a single compact extra dimension in pure gravity in which case the transition is first order~\cite{Gubser:2001ac}. Recently this calculation was generalized by Sorkin to black strings in arbitrary number of extended dimensions. He found that the phase transition becomes second order in more than twelve extended dimensions~\cite{Sorkin:2004qq}. It was also claimed that the critical number of dimensions, when there is a change in the phase transition order, depends on whether the phase transition happens at fixed black string mass (in microcanonical ensemble) or at fixed temperature (in canonical ensemble).  In the latter case the transition becomes of the second order in more then eleven extended dimensions~\cite{Kudoh:2005hf}. Later Kol and Sorkin considered the case with an arbitrary number $d$ of extended dimensions and an arbitrary number $p$ of the extra dimensions compactified on the torus $\mathbf{T}^p$~\cite{Kol}. In the special case when the sizes of all $p$ circles are equal they found that the phase transition order depends only on the number $d$ of extended dimensions and not on $p$. This result is explained by the fact that it is thermodynamically preferable for the GL instability to develop only along one of the circles on the torus. Therefore the toric black brane with $p > 1$ behaves effectively like the $p=1$ black string.

In this paper we determine the order of the phase transition for a black brane on a two-sphere compactification manifold. We use the  spontaneously compactified $\mathcal{M}_d \times \mathbf{S}^2$ solution of the Einstein-Maxwell theory in $D = d + 2$ dimensions as the background geometry~\cite{RandjbarDaemi}. The presence of the Maxwell field is necessary in order to have a non-flat compactification manifold. Aside from having a non-gravitational matter field this case has two important features in comparison to the flat compactification set-ups studied earlier which affect the properties of the phase transition. First, the two-sphere is not a direct product of two flat compact dimensions, and thus it does not support a mode of instability analogous to the modes along a single circle on the torus. Instability on $\mathbf{S}^2$ inevitably feels the presence of both compact dimensions and in this respect is more similar to the mode on the torus when the inhomogeneities along both circles are excited with equal amplitude (this mode is referred to as the ``diagonal" mode in~\cite{Kol}).
The other important difference is that the size of the two-sphere is fixed by the parameters of the theory. Therefore there is no rescaling freedom which in case of the flat compactifications accounts for the fact that the size of the compact dimension can be set arbitrarily. In the terms of dimensional reduction, the radion field is stabilized and has a mass comparable to the inverse radius of the compact two-sphere. Because of the  absence of an internal length scale for the flat compactification case, a set of thermodynamical quantities invariant under rescaling was introduced in~\cite{Gubser:2001ac} in order to study the phase transition. The results obtained in such a way correspond to the situation when the size of the compact extra dimension is held fixed and the radion is infinitely heavy. Since the critical black hole size is comparable to the radius of the two-sphere, the radion mass in our set-up is naturally of the same order as the phase transition temperature. Therefore one can expect the presence of the dynamical radion to play a non-trivial role in the phase transition. The presence of these features suggests that the phase transition order for the spherical black brane could be different from the case of the flat toric compactification.

In this paper we follow the method described in~\cite{Kol} to determine the order of the phase transition. The method employs the Landau-Ginzburg theory of phase transitions and is favoured in comparison to the original Gubser's computation since there is no need to compute any third order metric perturbations around the critical black brane. Considering the perturbations up to the second order is sufficient in order to compute the free energy and the entropy differences between the uniform and non-uniform black brane branches.  The signs of these differences define the phase transition order in canonical and microcanonical ensembles respectively. As a result we find that the transition for the spherical black brane in microcanonical ensemble is of the second order when the number of extended dimensions exceeds eleven. This coincides with the critical dimension for the ``diagonal'' mode on the two-torus found in~\cite{Kol}. In the canonical ensemble the difference of the free energy between the uniform and non-uniform black branes changes its sign when there are more than nine extended dimensions. It is lower than the corresponding critical dimension for the ``diagonal'' mode on the two-torus. However, since the canonical ensemble itself is ill-defined for the black holes in asymptotically flat extended dimensions due to their negative specific heat, it does not make sense to associate this critical behaviour of the free energy with the change of the phase transition order. The relation between the order of phase transition in the canonical and microcanonical ensembles for a generic system is discussed in the appendix~\ref{sec:mismatch}. Therefore we indeed found that the phase transition order for the spherical black brane does not coincide with the case of the flat toric compactification.

The rest of the paper is organized as follows. In section~\ref{sec:setup} we introduce the set-up and briefly describe how to apply the Landau-Ginzburg description of phase transitions to the black branes. We provide the details for the perturbative computation of inhomogeneous black brane solution in section~\ref{sec:computation}. The results for the phase transition order in various number of extended dimension are presented and discussed in section~\ref{sec:results}.

\section{Set-up and Ginzburg-Landau method}\label{sec:setup}

One of the simplest non-flat compactifications with a stabilized size of the extra dimensions is the spontaneous compactification solution of the 6D Einstein-Maxwell theory described in a great detail in~\cite{RandjbarDaemi}. In this solution the Maxwell vector field has a magnetic monopole configuration on an external two-sphere. It is possible to fine-tune the six-dimensional cosmological constant with respect to the magnetic flux so that the four-dimensional cosmological constant is zero and the remaining four dimensions are flat. It is straightforward to generalize this solution to $d$ extended dimensions. The fine-tuning condition between the cosmological constant, the vector field coupling, and the Newton's constant in $D \equiv d+2$  dimensions remains the same as in $d = 4$ case, i.e.
\be\label{eq:cc}
\Lambda_D = - \frac{e^2}{4\pi\, G_D}\;.
\ee

The phase transition takes place for a uniform black brane which looks like a spherically symmetric Schwarzschild black hole in the $d$ extended dimensions and completely wraps the external two-sphere. The Euclidean line element is given by
\be\label{eq:bkg}
ds^2 = f(r) dt^2 + \frac1{f(r)} dr^2 +r^2 d\Omega^2_{d-2} + a^2 \l d\theta^2 + \sin^2\theta\, d\phi^2\r \;,
\ee
where
\be
f(r) = 1 - \l\frac{r_0}{r}\r^{d-3}\;.
\ee
The first three terms in~\eqref{eq:bkg} correspond to a Schwarzschild-Tangherlini black hole in $d$-dimensional flat space, and the last two terms correspond to the two-sphere of the compact extra dimensions. The radius $a$ of this sphere is determined by the cosmological constant scale:
\be\label{eq:radius}
a^2 = - \frac1{2\,\Lambda_D}\;.
\ee
The mass and the inverse temperature of such a black brane are related to the horizon radius $r_0$ by the standard relations for a $d$-dimensional Schwarzschild black hole
\be\label{eq:temp}
M = \frac{(d-2)\, \Omega_{d-2}}{16\pi\,G_d}\, r_0^{d-3},\qquad \beta = \frac{4\pi}{d-3}\,r_0\;.
\ee
Here $G_d = G_D / (4\pi a^2)$ is the $d$-dimensional Newton's constant and $\Omega_{d-2}$ is the area of the $(d-2)$-dimensional unit sphere.

In the Landau-Ginzburg approach in order to study the order of phase transition in canonical ensemble it is sufficient to know the local behaviour of the free energy of the system around the critical point. A detailed description of this method in the context of black hole phase transitions is given in the reference~\cite{Kol} which we follow closely in our calculation. One first computes the local expansion of the free energy as a function of the order parameter and the temperature which in our case plays the role of the parameter that controls the onset of the transition. The role of the order parameter is played by the amplitude $\lambda$ of the inhomogeneous perturbations in the metric and vector field. It shows the degree of the deviation of the black brane from the uniform solution and is also our perturbative expansion parameter. In order to compute other thermodynamic characteristics it is useful to know the free energy as the function of the inverse temperature. We expand the free energy up to the fourth order in $\lam$ around critical point:\footnote{We checked that the odd terms in $\lam$ do not appear in the expansion of the black brane free energy.}
\be\label{eq:free}
F(\lambda; \beta) \simeq F_0(\beta) + \scA \l\frac{\db}{\beta_*}\r \lam^2 + \mathcal{C} \,\lam^4 \;,
\ee
where $F_0(\beta) = r_0^{d-3}$ is the free energy for the uniform unperturbed black brane\footnote{We follow~\cite{Kol} and omit in thermodynamic potentials the constant factor of $\frac{\Omega_{d-2}}{16\pi\,G_d}$ coming from the GR action.}, $\beta_*$ is the inverse critical temperature, and $\db \equiv \beta - \beta_*$. The values of the coefficients $\scA$ and $\scC$ in the expansion~\eqref{eq:free} define the local thermodynamics completely. The phase transition occurs when the black brane becomes smaller than a certain critical size, which means that $\db$ becomes negative. This fixes the sign of $\scA$ to be positive in order for the uniform phase $\lam =0$ to become an unstable extremum of the free energy for $\db<0$. In the case when $\scC$ is positive there is a minimum of the free energy for $\db<0$ located at
\be
\lam_*^2 \equiv - \frac{\mathcal{A}}{2 \scC} \l\frac{\db}{\beta_*}\r\;.
\ee
The presence of the non-trivial minimum in the vicinity of the uniform phase signals a smooth second order phase transition towards the slightly non-uniform phase with $\lam = \lam_*$.
The difference in free energies between the non-uniform and uniform black branes is of the fourth order in perturbative expansion parameter $\lam$ and is given by
\be\label{eq:deltaF}
F_*(\beta) - F_0(\beta) \simeq - \frac{\scA^2}{4\,\scC} \l\frac{\db}{\beta_*}\r^2 = - \,\scC\,\lam_*^4 \;.
\ee

The mass of the black brane can be obtained from the free energy using the relation
\be
M(\lambda; \beta) = \frac{\d (\beta\,F(\lambda; \beta))}{\d \beta}\;.
\ee
The entropy $S(\lambda; M)$ can be computed by performing a Legendre transform of $\beta F(\lambda; \beta)$ with respect to $\beta$ and is given by
\be
S(\lambda; M) \simeq S_0(M) - \beta_* \frac{\scA}{(d-3)}\,\l\frac{\delta M}{M_*}\r \lam^2 - \beta_* \l \scC - \frac{\scA^2}{2(d-2)(d-3)} \r\,\lam^4\;.
\ee
Here $S_0(M) = 4\pi\,r_0^{d-2}$ is the uniform black brane entropy, $M_*$ is the critical mass, and $\delta M \equiv M - M_*$. The entropy as a function of mass determines the behaviour of the system in microcanonical ensemble. In full analogy with the canonical ensemble case the order of the phase transition in the microcanonical ensemble is determined by the sign of the coefficient in front of the fourth order in $\lambda$ term, which also gives the sign of the difference between the entropies of the non-uniform and uniform phases (cf.~\eqref{eq:deltaF}):
\be\label{eq:ds}
\frac{S_* - S_0}{S_0} \simeq \frac1{d-3} \l \scC - \frac{\scA^2}{2(d-2)(d-3)} \r \, \lambda_*^4 \equiv \sigma_2\,\lam_*^4\;.
\ee
If the non-uniform branch has higher entropy than the uniform one, i.e. when $\sigma_2 > 0$, the phase transition in microcanonical ensemble is of the second order, and black brane settles in the stable non-uniform branch.

The free energy of the black brane as a function of the metric and the vector field potential is given by the Euclidean action of the Einstein-Maxwell theory evaluated on the corresponding solution:
\be\label{eq:action}
\beta F = I_E[g_{\mu\nu},V_\mu] \equiv - \frac1{8\pi G_D} \l \int_{\mathcal{M}} \half R + \int_{\mathcal{\d M}} \left[K - K^0\right] \r + \int_{\mathcal{M}} \l\frac14 F_{\mu\nu}^2 - \Lambda_D\r\;.
\ee
Here $R$ is the Ricci scalar, $K$ is the extrinsic curvature on the boundary at infinity and $F_{\mu\nu}$ is the vector field strength. The surface term corresponding to a reference geometry with the extrinsic curvature $K^0$ has to be subtracted in order to make the resulting free energy finite. In our case the reference geometry is $\mathbf{S}^1_\beta \times \mathbb{R}^{d-1} \times \sph_a$ with the Euclidean time period given by the inverse temperature $\beta$ and a fixed radius $a$ of the external two-sphere. The configuration space is spanned by the Euclidean solutions for $g_{\mu\nu}$ and $V_\mu$ that asymptote to this reference geometry.

The non-uniform solution can be found perturbatively by expanding the equations of motion and field deviations above the background in the powers of perturbative parameter $\lam$. The parameter $\lam$ also plays the role of the order parameter in the free energy expansion~\eqref{eq:free}. In order to compute the free energy up to the fourth order in $\lam$ it is sufficient to find the solution up to the second order. We collectively denote the metric and vector field deviations $\delta g_{\mu\nu} \equiv g_{\mu\nu} - g^{(0)}_{\mu\nu}$ and $\delta V_\mu \equiv V_\mu - V_\mu^{(0)}$  by $X \equiv \{\delta g_{\mu\nu}, \delta V_\mu\}$ and expand them in powers of $\lam$:
\be
X = \lam\,X^{(1)} + \lam^2\,X^{(2)} + \dots\;.
\ee
The first order perturbation $X^{(1)}$ is nothing else but the static inhomogeneous Gregory-Laflamme mode, and the second order perturbation $X^{(2)}$ corresponds to the back-reaction of the black brane on the GL mode.

In practice one expands the free energy in powers of $X$ and then plugs the solution for $X^{(1)}$ and $X^{(2)}$. For determining the order of the phase transition one is interested in the $\mathcal{O}(\lam^4)$ term in the free energy. Using the equations of motion one arrives to the following expression for the quartic coefficient $\scC$~\cite{Kol}:
\be\label{eq:c}
\scC = F_4[X^{(1)}] - F_2[X^{(2)}]\;.
\ee
Here the first term is the quartic in the field deviations $X$ term of the free energy expansion, which is evaluated on the first order solution $X^{(1)}$. The second term $F_2[X^{(2)}]$ is the quadratic in $X$ term evaluated on the second order perturbation $X^{(2)}$.

\section{Perturbative solution}\label{sec:computation}

In this section we are looking for a slightly inhomogeneous black brane solution as a perturbation around the homogeneous black brane~\eqref{eq:bkg}. The most general ansatz for Euclidean metric and vector field, which are static and spherically symmetric in $d$ extended dimensions reads
\begin{align}
ds^2 &= e^{2 A} f(r)\,dt^2 + \frac{e^{2 B}}{f(r)}\,dr^2 + e^{2 C}\,r^2\,d\Omega_{d-2}^2 +\nonumber \\\label{eq:ansatz1}
& + 2\,a^2\, G\,dr\,d\theta + a^2\, e^{2 H} \l e^{2 J} d\theta^2 +e^{-2 J} \sin^2\theta \,d\phi^2 \r\;, \\
V &= \frac1{2 e} \cos\theta\,d\phi - a\,L\,\sin\theta \, d\phi\;.\label{eq:ansatz2}
\end{align}
Here the functions $\{A, B, C, G, H, J, L\} = X$ parametrize the metric and vector field deviations and depend only on the $d$-dimensional radial coordinate $r$ and the angles on the external two-sphere $\theta$ and $\phi$. One could, in principle, also include the components of the metric, which are odd under the inversion on the two-sphere:
\be\nonumber
2\,a^2\, G_{odd}\,\sin\theta\,dr\,d\phi + 2\,a^2 H_{odd}\,d\theta\,d\phi\;,
\ee
as well as the even under the inversion component of the vector field (the monopole background itself is odd):
\be\nonumber
a\,L_{even}\,d\theta\;.
\ee
However it is always possible to choose a gauge in which these components are not excited and, therefore, we omit them. It is useful to expand the set of functions $X$ in spherical harmonics on the external two-sphere. It is always possible to choose the direction of the inhomogeneous mode to contain only the harmonics with $m = 0$ that do not depend on $\phi$. The spherical harmonics analysis also constrains considerably the possible $\theta$-dependence: the first order mode $X^{(1)}$ contains only the first $l =1$ harmonic, whereas the second order mode $X^{(2)}$ contains only the $l=0$ and $l = 2$ harmonics. Moreover all the functions $X$ can be subdivided according to their transformation properties under the coordinate transformations on the two-sphere into scalar, vector and tensor ones. The explicit expansion for the scalar quantities $X^{s} = \{A, B, C, H\}$ reads
\be\label{eq:scalar}
X^{s}(r,\theta) = \lam\, x_1^{s}(r)\cdot\cos\theta + \lam^2 \l x_0^{s}(r) + x_2^{s}(r)\cdot\half\,\l3\cos^2\theta - 1\r\r \;,
\ee
and the vector quantities $X^v = \{G, L\}$ are expanded as
\be\label{eq:vector}
X^v(r,\theta) = - \lam\, x_1^v(r)\cdot\sin\theta - \lam^2\, x_2^v(r)\cdot6\cos\theta \sin\theta \;.
\ee
The tensor harmonic parametrized by $J$ appears only at $l=2$ and is given by
\be
J(r,\theta) = \lam^2 \, j_2(r)\cdot\frac32\,\sin^2 \theta\;.
\ee
Hence in this ansatz all the metric and vector field perturbations are parametrized by the set of functions $x_i^{s} \equiv \{ a_i, b_i, c_i, h_i\}$, $x_i^{v} \equiv \{ g_i, l_i\}$, and $j_2$ of a single variable $r$.

The ansatz~\eqref{eq:ansatz1},\eqref{eq:ansatz2} does not fix the coordinate transformation redundancy in the $(r,\theta)$ plane and two more gauge fixing conditions have to be specified. These conditions can be specified independently at each order of perturbation theory and for each spherical harmonic label $l$. We use this freedom later in order to simplify the resulting equations.

The non-uniform black brane we are looking for has to satisfy the Euclidean equations of motion which in our case read
\begin{align}
R_{\mu\nu} &= 8\pi \,G_D \l F_{\mu\lam}{F_\nu}^\lam - \frac1{2\,d}\, g_{\mu\nu}\, F^2 \r - \frac2d \,g_{\mu\nu}\, \Lambda_D\;,\\
\nabla^\mu F_{\mu\nu} & = 0\;.
\end{align}
After plugging in the ansatz~\eqref{eq:ansatz1},\eqref{eq:ansatz2} and expanding up to the first order in $\lam$ one obtains a set of ordinary linear differential equations for the functions $x^s_1(r)$ and $x^v_1(r)$. After substituting also the background solution conditions~\eqref{eq:cc} and~\eqref{eq:radius} for the vector coupling and the cosmological constant, the remaining parameters in the resulting equations are the radius of the external sphere $a$ and the black brane horizon size $r_0$. For simplicity we set $r_0$ to be the unit of length. Then the only parameter in the equations is the dimensionless value of the radius $a$ in $r_0$ units. A regular solution exists only for a particular value of $a$ and defines the static Gregory-Laflamme mode on the black brane.


In linear order we use the particular choice of the gauge fixing condition proposed in~\cite{Kol}:
\begin{align}
b_1 &= \frac{r f' \, a_1 + 2(d-2)f\,c_1}{r f' + 2(d-2)f}\;, & l_1 &= 0\;.\label{eq:gauge}
\end{align}
In this gauge all linear equations are reduced to a single second order equation for $c_1(r)$:
\be\label{eq:master}
\frac1{r^{d-2}}\, \l r^{d-2} f \,c_1'\r' + \frac{2(d-1)(d-3)f'^2}{\l2(d-2)f+r f'\r^2} \,c_1 = \frac2{a^2}\,c_1\;.
\ee
Two conditions have to be imposed on the function $c_1$ in order to specify the solution. One of them corresponds to the choice of the normalization of $c_1$. We fix it by imposing the condition $c_1(1) =1$ adopted in previous works on black string~\cite{Gubser:2001ac,Sorkin:2004qq,Kudoh:2005hf,Kol}. The other condition arises if one requests the solution to be regular at the horizon $r = 1$ and is given by
\be
\frac{c_1'(1)}{c_1(1)} = \frac{2}{a^2 (d-3)}-(2 d-2) \;.
\ee
However for generic values of the parameter $a$ this regular solution grows exponentially for large $r$. Therefore one is left with a one-parameter shooting problem in which the value of $a$ is adjusted so that $c_1$ decays at large $r$. The equation~\eqref{eq:master} coincides with the first order equation (3.6) of~\cite{Kol} for the GL mode in the case of the black string and the black brane in the toric compactification. The critical size of the sphere $a_{GL}$ at which the GL instability sets in is related to the critical wavelength of the GL mode on the black string as:
\be
k_{GL}^2 = 2/{a_{GL}^2}\;.
\ee

In the gauge~\eqref{eq:gauge} all other components of the metric can be explicitly expressed in terms of $c_1(r)$ as:
\begin{align}
a_1 & = - (d-2)\, c_1\;,\\
g_1 & = - \frac{(d-2) \left(r \,f' - 2 \,f \right)} {2 r \,f' + 4 (d-2) \,f} \, c_1'  - \frac1{r \left( r \,f' + 2 (d-2) \,f \right)^2} \, (d-2) \Big{(} (d-3) r^2 \,f'^2 + \notag\\
& + r (((d-10) d+17) \,f+2 (d-3)) \,f' -  4 (d-3) (d-2) (\,f-1) \,f \Big{)} \, c_1\;,\\
h_1 & = 0\;.
\end{align}

At the second order the perturbations contain $l=0$ and $l=2$ modes which can be treated independently. The second order equations are  inhomogeneous linear ODE's for $x_0$, $x_2$ and $j_2$ that contain source terms quadratic in the first order perturbations found above. We start by considering the zero modes $a_0, b_0, c_0$ and $h_0$. The function $h_0$ can be separated from the other zero modes and is defined  by the equation
\be\label{eq:h0}
\frac1{r^{d-2}}\, \l r^{d-2} f \,h_0'\r' - \frac{2(d-2)}{d\,a_{GL}^2}\,h_0 = S_{h_0}[b_1,c_1,g_1]\;.
\ee
The exact form of the source term $S_{h_0}[b_1,c_1,g_1]$ is given by~\eqref{eq:srch0}. The regularity condition on the horizon fixes the value of the first derivative $h_0'(1)$ in terms of the value of $h_0(1)$, while the latter is adjusted in order to find the solution that decays at large $r$. 

Instead of finding the solution for $a_0, b_0$ and $c_0$ it is more efficient not to fix the gauge at all but to rewrite the equations in terms of the gauge-invariant combinations defined as:
\begin{align}
u &\equiv a_0 + b_0 - (r\,c_0)'\;,\\
w &\equiv c_0 - \frac{2f}{r\,f'}\,a_0\;.
\end{align}
The free energy is obviously a gauge invariant quantity and can be expressed in terms of $u$ and $w$. Thus there is no need to fix any particular gauge and determine $a_0, b_0$ and $c_0$.

The equation for $u$ is a first order differential equation
\be\label{eq:u}
u' = \frac{2}{d}\,r\,h_0'' + S_{u}\;,
\ee
with the source term given in~\eqref{eq:srcu}. This equation can be straightforwardly integrated numerically. For the metric to match the flat reference geometry the constant of integration $u(1)$ should be chosen so that $u$ vanishes at infinity. The equation for $w$ also happens to be of the first order:
\be\label{eq:w}
w' + \frac2{r^{d-4}}\,u +\frac2{d-2}\frac{f}{f'}\,h_0'' - \frac{2f}{r f'}\,h_0' -\frac4{d \, a^2f'} h_0 = S_w\;.
\ee
with the source given in~\eqref{eq:srcw}.  The equation can be integrated once the solution for $u$ is known. The constant of integration $w(1)$ can be fixed by considering the temperature of the perturbed black brane. More precisely the deviation of the temperature from the critical one can be expressed in terms of the metric perturbations as
\be\label{eq:wbeta}
\frac{\db}{\beta_*} = b_0(1) - a_0 (1) = u(1) + w(1) + w'(1)\;.
\ee
Thus different values of $w(1)$ correspond to black branes with different temperatures. However, in order to obtain the free energy of the black brane at the critical point we consider only solutions with the critical temperature. The integration constant $w(1)$ is then fixed by demanding the $\db$ to be zero. The freedom in the choice of $w(1)$ is related to the fact that among the linear perturbations around any black hole there always exists a mode corresponding to the infinitesimal change of the size of the black hole.

The equations for the $l=2$ spherical harmonic components can also be separated in two groups. First we solve for the variables $h_2, j_2$, and $l_2$, the equations for which are independent from the other variables. We adopt the same gauge condition $l_2 = 0$ as in the linear order. The system of equations for the $h_2$ and $j_2$ then takes the form:
\begin{align}
\frac1{r^{d-2}}\, \l r^{d-2} f \,h_2'\r' & = \frac2{a^2} \l \frac{7d - 2}{d} h_2 + 6 j_2\r + S_{h_2}\;,\label{eq:h2}\\
\frac1{r^{d-2}}\, \l r^{d-2} f \,j_2'\r' & = - \frac2{a^2}\,h_2 + S_{j_2}\;,\label{eq:j2}
\end{align}
with the source terms given by equations~\eqref{eq:srch2} and~\eqref{eq:srcj2}. It is straightforward to bring this system to a solvable from by taking appropriate linear combinations of $h_2$ and $j_2$. After that the equations for these linear combinations can be solved separately. In order to obtain solution one, as before, has to impose the regularity condition on the horizon and solve the one-parameter shooting problem by adjusting the values of the functions on the horizon in order to obtain the solution that decays at infinity.

In analogy to the linear order one can fix the gauge in such a way that it would be possible to express the remaining variables $a_2, b_2$, and $g_2$ algebraically in terms of $c_2$, which obeys a single second order ODE. The gauge fixing condition~\eqref{eq:gauge} is modified at the second order by the presence of a term quadratic in the first order variables:
\be
b_2 = \frac{r f'a_2 + 2(d-2)f\,c_2}{r f' + 2(d-2)f} - \frac43\frac{r f}{r f' + 2(d-2f)} \,b_1\,g_1\;.
\ee
The equation for $c_2$ is similar to the equation~\eqref{eq:master} that defines the first order perturbation:
\begin{multline}\label{eq:c2}
\frac1{r^{d-2}}\, \l r^{d-2} f \,c_2'\r' + \frac{2(d-1)(d-3)f'^2}{\l r f' + 2(d-2)f \r^2} \,c_2 - \frac6{a^2}\,c_2 + \\
 +2  \frac{ \l r f' + (d-3)f \r f'^2}{\l r f' + 2(d-2)f \r^2} \,\l h_2 + 2 j_2 \r + \frac4{d\,a^2}\, h_2 + S_{c_2} = 0\;,
\end{multline}
with the source term given by~\eqref{eq:srcc2}. Thus finding the solution in this sector boils down to another one-parameter shooting problem for the value $c_2(1)$.
After finding the solution for $c_2$ one can determine the remaining variables $a_2$ and $g_2$ as:
\begin{align}
a_2 &= -(d-2) c_2 - h_2 - 2 j_2 - \frac{(d-1)(d-2)}6\,c_1^2 - \frac{a^2}{6}\,f\,g_1^2 \;,\\
g_2 &= - \frac1{6} \frac{(d-2) (r f'- 2 f)}{r f' + 2(d-2) f} \, c_2 - \frac1{12}\,b_2'
- \frac1{12} \frac{3 r f'- 2(d-2) f}{r f' + 2(d-2) f} \, h_2 - \notag\\
& - \frac1{6} \frac{r f'- 2(d-2) f}{r f' + 2(d-2) f} \, j_2 + S_{g_2}\;. \label{eq:g2}
\end{align}
The explicit expression for the source $S_{g_2}$ can be found in~\eqref{eq:srcg2}.

\section{Results and discussion}\label{sec:results}

Having found the solution for the metric and the vector field we can compute the coefficients in the free energy expansion~\eqref{eq:free}. By substituting the solution with $\db=0$ in the Euclidean action~\eqref{eq:action} one obtains an expression for the free energy quartic in $\lam$, from which the coefficient $\scC$ can be determined. Alternatively one can use the expression~\eqref{eq:c}, which is simpler to evaluate numerically. We found the numerical mismatch between the two different expressions for $\scC$ to be less than a percent. By changing the parameters of numerical integration we found the change in the values of $\scC$ to be at the same level, which thus can serve as the estimate of the numerical error. The values of $\scC$ for a various number of extended dimensions $d$ are listed in table~\ref{tab}. Note the change of sign of the quartic coefficient for $d > 9$. For a thermodynamically stable system it would mean that the phase transition for $d > 9$ is of the second order in canonical ensemble. 
\begin{table}[t]
\centering
\begin{tabular}{ c c c c c c c c c c c}
\hline\hline
$d$	& 4	& 5	& 6	& 7	& 8	& 9	& 10	& 11	& 12	& 13 \\
\hline
$\scA$		&0.544	& 1.73	& 3.59	& 6.15	& 9.39	& 13.3	& 17.9	& 23.2	& 29.2	& 35.8 \\
$\scC$		& -0.114	& -0.384	& -0.785	& -1.20	& -1.41	& -1.10	& 0.14	& 2.84	& 7.62	& 15.2 \\
$\sigma_2$	& -0.188	& -0.317	& -0.441	& -0.536	& -0.576	& -0.535	& -0.389	& -0.112	& 0.322	& 0.938 \\\hline
\end{tabular}
\caption{The coefficients $\scA$ and $\scC$ in the free energy expansion~\eqref{eq:free} and the entropy variation $\sigma_2$ defined in~\eqref{eq:ds} for a different number of extended dimensions $d$. The change of the sign of $\sigma_2$ between $d = 11$ and 12 indicates that the phase transition in microcanonical ensemble becomes of the  second order for $d > 11$. The analogous change in the free energy behaviour happens between $d = 9$ and 10. \label{tab} }
\end{table}

\begin{figure}[t!]
\centering
\includegraphics[width=.7\textwidth]{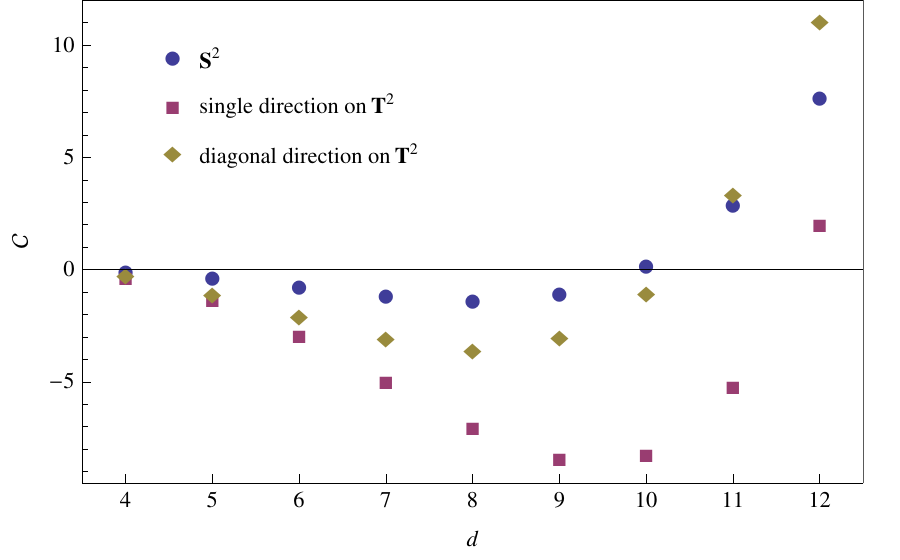}
\caption{The quartic coefficient $\scC$ in the free energy expansion~\eqref{eq:free} for the spherical black brane (circles) in a various number of extended dimensions $d$ in comparison to the single direction (squares) and ``diagonal" (dimonds) modes of the black brane on a square two-torus from~\cite{Kol}.\label{fig:C}}
\end{figure}

It is instructive to compare the obtained behaviour of $\scC$ for a black brane on a two-sphere with the case of a flat compactification on the square two-torus $\mathbf{T}^2$ considered by Kol and Sorkin in~\cite{Kol}. In the latter case there are two independent inhomogeneous modes corresponding to the two circles of $\mathbf{T}^2$. In order to study the free energy one can consider two limiting cases: when only a single mode along one of the two circles is excited, or when both of the modes are excited with equal amplitude, the so-called ``diagonal'' mode~\cite{Kol}. The quartic coefficient $\scC$ in all three cases is presented in dependence of the number of extended dimensions $d$ in figure~\ref{fig:C}. We see that the single direction mode on a torus has lower free energy and thus thermodynamically favourable. Due to this fact the toric black branes during the phase transition effectively behave like black strings, with only the mode along a single circle being excited. In contrast, on the spherical black brane there is only one mode, and its dependence on the number of extended dimensions $d$ is different from the modes on $\mathbf{T}^2$. The change of the sign of the coefficient $\scC$ for the spherical black brane happens between $d = 9$ and 10.

The behaviour of the black brane in microcanonical ensemble is determined by the sign of the coefficient $\sigma_2$ in the difference of the entropy between the non-uniform and uniform black branes~\eqref{eq:ds}. In order to find $\sigma_2$ the quadratic coefficient $\scA$ should be determined. There are two independent methods how to compute $\scA$. First, one can take a first variation of the free energy~\eqref{eq:action} with respect to the temperature, which at the leading order in $\lam$ can be expressed using only the solution with $\db = 0$. This method was applied in~\cite{Kol} and gives
\be
\scA = \frac43 \frac{(d-1)(d-2)}{a^2} \int_1^\infty c_1^2\,r^{d-2} \, dr \;.
\ee
Alternatively one can use the solution of~\eqref{eq:w} with $\db \neq 0$, i.e. keep the integration constant for the zero harmonic $w(1)$ initially unspecified. The temperature dependence of the free energy~\eqref{eq:action} is then obtained by using the relation~\eqref{eq:wbeta} relating $\db$ to $w(1)$. The coefficient $\scA$ is given by
\be
\scA = (d-1)(d-2)\,w(r \to \infty)\;,
\ee
where the asymptotic value $w(r \to \infty)$ is taken from the solution with $\db = 0$.

\begin{figure}[t]
\centering
\includegraphics[width=.7\textwidth]{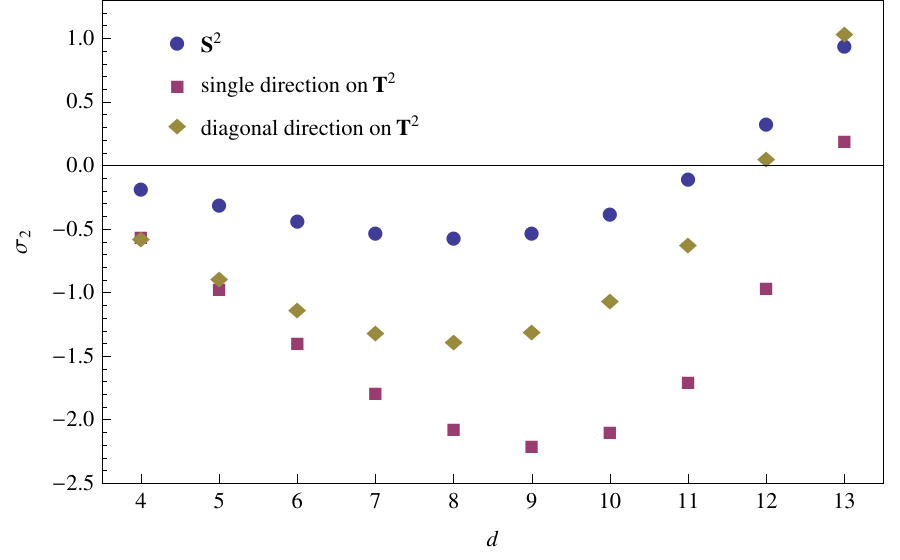}
\caption{The coefficient $\sigma_2$ of the entropy difference between the non-uniform and uniform spherical black branes (diamonds) in comparison to the single direction (circles) and ``diagonal" (squares) modes of the toric black brane from~\cite{Kol}.\label{fig:dS}}
\end{figure}

The resulting values of $\sigma_2$ in dependence on the number of extended dimensions $d$ are given in the table~\ref{tab} and presented in comparison to the case of the toric black brane in figure~\ref{fig:dS}. For $d > 11$ the non-uniform black brane has larger entropy than the uniform one, and the phase transition in microcanonical becomes of the second order. We note that in microcanonical ensemble the critical number of extended dimensions for the spherical black brane coincides with the one for the diagonal inhomogeneous mode of the toric black brane.

For $d=10,11$ the signs of the quartic coefficients in the free energy and entropy, and consequently the predicted orders of phase transition in canonical and microcanonical ensembles, are different. The possibility of such a situation can be seen already from the equation~\eqref{eq:ds} for the entropy difference, since the entropy difference can remain negative even if the coefficient $\scC$ would turn to be positive. The details of this effect are discussed in appendix~\ref{sec:mismatch}.

\acknowledgments

The author is indebted to Sergey Sibiryakov, Gia Dvali, Dima Levkov and Valery Rubakov for fruitful discussions and L\={a}sma Alberte for careful reading of the draft. The research was supported by Alexander von Humboldt Foundation.

\appendix

\section{The origin of the phase transition order mismatch in canonical and microcanonical ensembles}\label{sec:mismatch}

We consider a generic form of the free energy expansion in the vicinity of the critical temperature $\beta_*$
\be
F(\beta ,\lam) \simeq f_0 +f_1 \frac{\db}{\beta_*} + f_2 \l\frac{\db}{\beta_*}\r^2 + \mathcal{A}\l\frac{\db}{\beta_*}\r \lam^2 + \mathcal{C} \,\lam^4 \;,
\ee
where $\lam$ is the order parameter, and $\db \equiv \beta - \beta_*$, cf.~\eqref{eq:free}.
The entropy is given by a Legendre transform of $\beta F$ with respect to $\beta$ and reads
\be
S(M,\lam) \simeq \beta_* f_1 + \beta_* \delta M + \beta_* \frac1{4(f_1+f_2)} \delta M^2 - \frac{\beta_*\mathcal{A}}{2(f_1+f_2)} \delta M \lam^2 - \beta_* \left[ \mathcal{C} - \frac{\mathcal{A}^2}{4(f_1+f_2)} \right] \lam^4 \;,
\ee
with $\delta M \equiv M - M_*$. The specific heat $c_u$ of the uniform black brane with $\lam =0$ kept fixed is given by
\be
c_u = - \beta \l\frac{\d^2 (\beta F)}{\d\beta^2}\r_{\lam=0} \simeq -2 \beta_* (f_1+f_2)\;.
\ee
By using this expression the entropy expansion can be rewritten in the following form:
\be\label{eq:ent}
S(M,\lam) \simeq \beta_* f_1 + \beta_* \delta M - \frac{\beta_*^2}{2 c_u} \delta M^2 - \frac{\beta_*^2\mathcal{A}}{c_u} \delta M \lam^2 - \beta_* \left[ \mathcal{C} + \frac{\beta_*\mathcal{A}^2}{2c_u} \right] \lam^4\;.
\ee
If the specific heat $c_u$ is negative, the coefficient $\tilde\scC \equiv \left[ \mathcal{C} + \frac{\beta_*\mathcal{A}^2}{2c_u} \right]$, which determines the stability of the non-uniform branch in microcanonical ensemble, may become negative even for positive values of $\scC$.

In the case of a positive $\scC$ a non-trivial minimum of the free energy, located at
\be
\lam_*^2 = - \frac{\mathcal{A}}{2 \scC} \l\frac{\db}{\beta_*}\r\;,
\ee
appears for $\db < 0$. Therefore, in canonical ensemble the system should settle in the non-uniform phase with $\lam = \lam_*$. The specific heat in this non-uniform phase is given by
\be
c_{nu} = - \beta \frac{\d^2 \beta F(\beta,\lam_*(\beta))}{\d\beta^2} \simeq c_u + \frac{\beta_*\mathcal{A}^2}{2\scC}\;.
\ee
Comparison with \eqref{eq:ent} gives the following relationship between the quartic coefficients in the two ensembles and the specific heats in the different phases:
\be\label{eq:ratio}
\frac{\tilde\scC}{\scC} = \frac{c_{nu}}{c_u}\;.
\ee
This condition holds for any thermodynamic system and tells that the signs of the quartic coefficients in the free energy and the entropy expansions are different if and only if the specific heat has different sign in different phases. Hence, the mismatch between the phase transition orders in canonical and microcanonical ensembles can happen only if the specific heat changes its sign in the course of transition from the uniform to the non-uniform phase. Such a situation is indeed observed in some gravitational systems (c.f.~\cite{LyndenBell} and references therein). If the system is thermodynamically stable before the phase transition, i.e $c_u > 0$, then a situation is possible when the non-uniform branch is stable in the microcanonical ensemble ($\tilde\scC > 0$) and unstable in the canonical ensemble ($\scC < 0$). In such a case the system exhibits a second order phase transition in the microcanonical ensemble. In canonical ensemble the transition is of the first order and proceeds towards some third phase which is thermodynamically stable. This behaviour is attributed to the fact that in this case the system has negative specific heat in the non-uniform phase, as can be seen from~\eqref{eq:ratio}.

The opposite situation takes place for the spherical black brane in the cases $d = 10, 11$. Specific heat flips its sign from negative on the uniform branch to positive on the non-uniform. Nevertheless it does not mean that the branch of the non-uniform branes is thermodynamically stable. Due to the change of sign of the specific heat the ``stable'' non-uniform phase, which appears in canonical ensemble for $\db < 0$, corresponds to the black branes with the mass larger than $M_*$. Thus the mass of the brane does not cross the GL critical value in the course of this transition, and the would be new phase is related to the local minimum of the entropy which appears for $\delta M > 0$. Thus it seems that positive specific heat of non-uniform branch in canonical ensemble is spurious, and the phase transition in canonical ensemble never 
proceeds towards the found non-uniform branch. Moreover in the case at hand the change of sign of the specific heat during the transition would mean that slightly non-uniform spherical black branes with flat asymptotics are thermodynamically stable in certain number of extended dimensions which does not seem to be the case.

\section{The source terms for the second order perturbation equations}\label{sec:source}

For the sake of completeness we present here the full expressions for the sources in back-reaction equations. The source terms for the $l=0$ mode equations~\eqref{eq:h0},~\eqref{eq:u} and~\eqref{eq:w}:
\begin{align}
S_{h_0} &= \frac23\,\frac1{r^{d-2}}\,\l r^{d-2} f \, g_1\r'  b_1+ \frac13\,f\,g_1\, b_1' - \frac1{3a^2} \, b_1^2 - \frac{(d-1)(d-2)}{3a^2}\,c_1^2\;,\label{eq:srch0}\\
S_u &=  \frac{d+1}3\,r\,c_1'^2 
- \frac23\frac{r}{f}\frac1{d\, a^2}\,b_1^2 -\frac23\,\frac{r}{f}b_1 \l \frac1{r^{d-2}} \l r^{d-2} f c_1' \r' - \frac1{a^2} c_1 \r - \notag\\
&- \frac13\,r\,b_1' c_1' -\frac13\frac{r f'}{(d-2)\,f} b_1\,b_1' - \frac23\,b_1 g_1 -\frac23\frac{r}{f}\frac1{r^{d-2}} \l r^{d-2} f\,c_1\,g_1\r' + \frac13 a^2 \l f \,g_1^2\r \;,\label{eq:srcu}\\
S_w &= -\frac23 \frac{d-3}{r^2 f'} \, c_1^2 - \frac{d-1}{3}\frac{f}{f'} c_1'^2
- \frac23 \l \frac{1}{d\,f'} + \,f \frac{a^2}{r}\r \,g_1^2 
+\frac23 \frac{f}{(d-2) f'} \,g_1\,b_1' + \notag \\
&+\frac23 \frac1{f'} \l \frac{rf' + (d-3)f}{r^2} + \frac1{a^2(d-2)} \r \, b_1^2 + \frac13 \frac{r f' +2(d-2)f}{(d-2) r f'}\,b_1\,b_1'\;.\label{eq:srcw}
\end{align}
The source term for the $c_2$ equation~\eqref{eq:c2} reads 
\begin{align}
S_{c_2} & = \frac1{3 a^2 \left(2 (d-2) \,f+r \,f'\right)^3} \, \Big{(} r^2 \,f'^3 \left(5 a^2 (d-2) (d-1) \,f'+4 (d-3) r\right) + \notag\\
& + 2 (d-2) \,f^2 \,f' \left(a^2 (d-6) (d-3) (d-1) \,f'-4 (d (5 d-13)+3) r\right) + \notag\\
& + (d-2) r \,f \,f'^2 \left(a^2 (d-1) (7 d-27) \,f'-2 (9 d+8) r\right)-8 (d-2)^2 (2 d (2 d-5)+5) \,f^3 \Big{)} \,c_1^2 + \notag\\
& + \frac{2 (d-2) \,f \left((d-3) \,f+r \,f'\right)}{6 (d-2) \,f+3 r \,f'} \, c_1'^2 + \frac{2 a^2 \,f^2  \left((d-2) \,f+r \,f'\right)}{3 (d-2) \left(2 (d-2) \,f+r \,f'\right)} \, g_1'^2 +\notag\\
& + \frac{2 \,f \left(2 (d-6) r \,f \,f'+4 (d-2) (d-1) \,f^2+5 r^2 \,f'^2\right)}{3 \left(2 (d-2) \,f+r \,f'\right)^2} \, c_1\,g_1' + \notag\\
& + \frac1{3 (d-2) d \left(2 (d-2) \,f+r \,f'\right)^2} \Big{(} 2 a^2 d r^2 \,f'^4 + 2 (d-2) r \,f \,f'^2 \left(4 a^2 d \,f'+r\right) + \notag\\
& + (d-2) \,f^2 \,f' \left(a^2 d (7 d-15) \,f'-4 (d+4) r\right)-16 (d-2)^2 (d+1) \,f^3\Big{)} \, g_1^2 + \notag\\
& + \frac{2 \,f \left((15 d-32) r \,f \,f'+2 (d-2) (8 d-15) \,f^2+5 r^2 \,f'^2\right)}{3 \left(2 (d-2) \,f+r \,f'\right)^2} \, c_1' \, g_1 + \notag\\
& + \frac1{3 r \left(2 (d-2) \,f+r \,f'\right)^2} 2\Big{(} (2 d-3) r^3 \,f'^3+2 (d-2) (4 d-7) r^2 \,f \,f'^2 + \notag\\
& + (d-2) (7 (d-5) d+40) r \,f^2 \,f'+2 (d-5) (d-2)^3 \,f^3\Big{)} \, c_1' \, c_1 + \notag\\
& + \frac{2  \,f\left((3 d-5) r \,f'+(d-5) (d-2) \,f\right)}{6 (d-2) \,f+3 r \,f'} \, c_1'' \, c_1 + \frac{2 a^2 \,f^2 \left((d-2) \,f+r \,f'\right)}{3 (d-2) \left(2 (d-2) \,f+r \,f'\right)} \, g_1'' \, g_1 \notag\\
& + \frac{2 a^2 \,f  \left(6 (d-2) r \,f \,f'+(d-2)^2 \,f^2+4 r^2 \,f'^2\right)}{3 (d-2) r \left(2 (d-2) \,f+r \,f'\right)} \, g_1' \, g_1 + \notag\\
& + \frac1{3 r \left(2 (d-2) \,f+r \,f'\right)^3} \, 2 \, \Big{(}3 (d-4) r^3 \,f \,f'^3+2 (d-1) (4 d-7) r^2 \,f^2 \,f'^2 + \notag\\
& + 2 (d-2) (d (9 d-26)+11) r \,f^3 \,f'+8 (d-2)^3 (d-1) \,f^4+3 r^4 \,f'^4\Big{)} \, c_1 \, g_1\;. \label{eq:srcc2}
\end{align}
The source terms for the rest of $l = 2$ mode equations~\eqref{eq:h2},~\eqref{eq:j2} and~\eqref{eq:g2}: 
\begin{align}
S_{h_2} &= \frac{(d-1)(d-2)}{3a^2}\,c_1^2 +\frac1{3a^2}\,b_1^2 +\frac{d-2}{3d}\,f\,g_1^2 -\frac23 \frac1{r^{d-2}} \l r^{d-2} f\,g_1\r'\,b_1 - \frac13\,f\,g_1\,b_1' \;,\label{eq:srch2}\\
S_{j_2} &= -\frac{(d-1)(d-2)}{3a^2}\,c_1^2 -\frac1{3a^2}\,b_1^2 -\frac13\,f\,g_1^2 +\frac23 \frac1{r^{d-2}} \l r^{d-2} f\,g_1\r'\,b_1 + \frac13\,f\,g_1\,b_1' \;,\label{eq:srcj2}\\
S_{g_2} &=- \frac{(d-1)(d-2)}{36}\frac{r}{r f' + 2(d-2)f} \l 2f\,c_1''\,c_1 - 2f\,(c_1')^2 + 3f'\,c_1'\,c_1\r - \notag \\
& - \frac1{r f' + 2(d-2)f}  \l \frac{7r}{18 a^2} \, b_1^2 +\frac{r}{18} \, f\,b_1'\,g_1 +  \frac{r}{3} \, f\,b_1\,g_1' + \frac{7r f' + 8(d-2) f}{18} \,b_1\,g_1 \r - \notag\\
& - \frac1{36}\frac{a^2}{r f' + 2(d-2)f} \Big{(}  r\,f^2\, (g_1^2)'' + \l 9 r\,f' + 4(d-2)\,f \r \,f\, g_1'\,g_1 + \notag\\
& + \frac12 \l 5 r\,f' + 2(d-2)\,f \r \,f'\, g_1^2 \Big{)} - \frac1{3}\frac{r f}{r f' + 2(d-2)f}\, g_1^2 \;.\label{eq:srcg2}
\end{align}

\end{document}